\documentclass[journal=jacsat,manuscript=article]{achemso}
\usepackage[version=3]{mhchem}
\usepackage{color}

\title{\ce{FePd2Te2}: an anisotropic two-dimensional ferromagnet with one-dimensional Fe-chains}
\author{Bingxian Shi,$^{1,6,\dag}$ Yanyan Geng,$^{1\dag}$ Hengning Wang,$^{2,3,\dag}$ Jianhui Yang,$^{4\dag}$ Chenglin Shang,$^{1,6}$ Manyu Wang,$^{1}$ Shuo Mi,$^{1}$ Jiale Huang,$^{1,6}$ Feihao Pan,$^{1,6}$ Xuejuan Gui,$^{1,6}$ Jinchen Wang,$^{1,6}$ Juanjuan Liu,$^{1,6}$ Daye Xu,$^{1,6}$ Hongxia Zhang,$^{1,6}$ Jianfei Qin,$^{5}$ Hongliang Wang,$^{5}$ Lijie Hao,$^{5}$ MingLiang Tian,$^{3*}$ Zhihai Cheng,$^{1*}$ Guolin Zheng$^{3*}$  and Peng Cheng$^{1,6}$}
\email{zhihaicheng@ruc.edu.cn,glzheng@hmfl.ac.cn,pcheng@ruc.edu.cn}

\affiliation[RUC]
{$^{1}$ Department of Physics, Key Laboratory of Quantum State Construction and Manipulation, Ministry of Education, Renmin University of China, Beijing 100872, China\\
	$^{2}$ Department of Physics, University of Science and Technology of China, Hefei 230031, Anhui, China\\
	$^{3}$ Anhui Key Laboratory of Low-Energy Quantum Materials and Devices, High Magnetic Field Laboratory, HFIPS, Chinese Academy of Sciences, Hefei, Anhui 230031, China\\
	$^{4}$ Quzhou University, Quzhou, Zhejiang 32400, China\\
	$^{5}$ China Institute of Atomic Energy, PO Box-275-30, Beijing 102413, China\\
	$^{6}$ Laboratory for Neutron Scattering, Department of Physics, Renmin University of China, Beijing 100872, China
}

\begin{document}
	
\begin{abstract}
	Two-dimensional (2D) magnets have attracted significant attentions in recent years due to their importance in the research on both fundamental physics and spintronic applications. Here, we report the discovery of a new ternary compound \ce{FePd2Te2}. It features a layered quasi-2D crystal structure with one-dimensional Fe zigzag chains extending along the $b$-axis in the cleavage plane. Single crystals of \ce{FePd2Te2} with centimeter-size could be grown. Density functional theory calculations, mechanical exfoliation and atomic force microscopy on these crystals reveal that they are 2D materials that can be thinned down to $\sim$5~nm. Magnetic characterization shows that \ce{FePd2Te2} is an easy-plane ferromagnet with $T_{c}\sim$183~K and strong in-plane uniaxial magnetic anisotropy. Magnetoresistance and anomalous Hall effect demonstrate that ferromagnetism could maintain in \ce{FePd2Te2} flakes with large coercivity. A crystal twinning effect is observed by scanning tunneling microscopy which makes the Fe chains right-angle bent in the cleavage plane and creates an intriguing spin texture. Besides, a large electronic specific heat coefficient up to $\gamma\sim${32.4 mJ/$mol^{-1}$$K^{-2}$} suggests \ce{FePd2Te2} is a strongly correlated metal. Our results show that \ce{FePd2Te2} is a correlated anisotropic 2D magnets that may attract multidisciplinary research interests.
\end{abstract}

\section{Introduction}

	Since the experimental demonstration of intrinsic ferromagnetism in monolayer \ce{CrI3}, \ce{Cr2Ge2Te6} and \ce{Fe3GeTe2}\cite{GongC,HuangB,JiangSW,DengYJ}, intense research interests in two-dimensional (2D) magnets has been spurred. Compounds of this type, either with van der Waals (vdW) or layered non-vdW structure, as long as they can be thinned down to a few atomic layers (typically less than 5~nm), are promising for developing next-generation ultrathin spintronic devices\cite{Burch,Rev_nonvdw}. Especially for 2D magnetic metals, their metallic nature may enable the interplay of both spin and charge degrees of freedom that lies at the heart of various spintronic architectures\cite{DengYJ}. Besides, 2D magnet is also an ideal platform for investigating how the Hamiltonians of the fundamental magnetism models (the Ising, XY and Heisenberg models) would behave in 2D limit\cite{Burch}. Therefore they are also important in fundamental physics research and exploring novel quantum effects. 
	
	The magnetic anisotropy is crucial for resisting thermal fluctuations and stabilizing long range magnetic order in 2D limit\cite{Burch}. Most 2D magnets studied to date have perpendicular magnetic anisotropy\cite{GongC,HuangB,JiangSW,DengYJ,May_Fe5,VI3,Fe3GaTe2,Fe5_John,CP_Fe5,Cr5Te8}, namely the easy-axis is $c$-axis which is perpendicular to the cleavage plane. Although large perpendicular magnetic anisotropy is beneficial for realizing high-density information storage devices, 2D magnets with an easy-plane magnetic anisotropy (moment lies in the cleavage plane) are also appealing for various spin-related research and applications. For example, \ce{CrCl3} is predicted to host 2D superfluid spin transport and topological magnetic textures due to its 2D-XY ferromagnetism\cite{CrCl3_science,CrCl3_nano,CrCl3_wang}. \ce{CrSBr} with the A-type antiferromagnetic structure and strong in-plane easy-axis anisotropy have shown large negative magnetoresistance coupled with magnetic order\cite{CrSBr1,CrSBr2}. Furthermore, by twisting two CrSBr ferromagnetic monolayers with an easy-axis in-plane spin anisotropy by 90$^{\circ}$, special field-induced features which are useful in controlling and addressing the magnetic information could be observed\cite{CrSBr_twist}. However, 2D magnets with in-plane anisotropy seem to be quite rare besides \ce{CrCl3} and \ce{CrSBr}. On the other hand, many 2D magnets which are currently under hot research may suffer from low magnetic transition temperature, sensitivity of air or lack of large single crystals for large-scale device integration. So finding new air-stable 2D magnets with high Curie temperature and peculiar magnetic properties may promote the development of this research field.
	
	In this work, we report the discovery of a new compound \ce{FePd2Te2}. It is air-stable and enables the growth of centimeters-large single crystal that can be exfoliated to few layers. By combined X-ray, magnetization, transport, specific heat, neutron scattering and scanning tunneling microscopy (STM) studies, \ce{FePd2Te2} is identified as a correlated hard 2D ferromagnet with Curie temperature of 183~K and strong in-plane easy-axis magnetic anisotropy. Besides, an intriguing spin texture is formed in \ce{FePd2Te2} due to crystal twinning effect.  \ce{FePd2Te2} opens up new possibilities in the research of 2D magnetism and spintronic applications.

\section{Experimental Section}

\subsection{Single Crystal Growth.}
	Single crystals of \ce{FePd2Te2} were grown by melting stoichiometric elements. Iron, palladium and tellurium powder were mixed and grounded in a molar ratio of 1:2:2. Deviation from this ratio would lead to the unsuccessful growth of large single crystals or target phase. Then the mixtures were placed in an alumina crucible and sealed in a quartz tube under vacuum conditions. The entire tube was heated in a box furnace to 800$\,^{\circ}\mathrm{C}$ and held for 2 days. Then it was cooled to 600$\,^{\circ}\mathrm{C}$ at a rate of 2$\,^{\circ}\mathrm{C}$/h followed by annealing at this temperature for 2 days before furnace-cooled to room temperature. In addition, we found that quenching the samples at 600$\,^{\circ}\mathrm{C}$ would improve the crystal quality as revealed from X-ray characterization. However, direct quenching in the initial growth process seems to break the large crystal into small pieces. By re-annealing and quenching the large single crystal grown by initial furnace-cooled method, one could obtain crystal with both large size and good quality.
	
\subsection{Characterization.}
    Single crystal X-ray diffraction (XRD) patterns were collected at 273 K using a Bruker D8 VENTURE PHOTO II diffractometer equipped with multilayer mirror monochromatized Mo K$\alpha$ ($\lambda=$0.71073 \AA) radiation. Unit cell refinement and data merging were performed using the SAINT program. An absorption correction was applied using Multi-Scans. Upon inspecting the systematic absences, it is clear that the structure belongs to the $P$ lattice. In XPrep, the suggested possible space groups are  $P2_1$ and $P2_1/m$ with the value of combined figure of merit (CFOM) 6.48 and 1.18, respectively. The mean value of $\mid E^{2}-1\mid$ (0.971) is closer to the value for a centrosymmetric space group (0.968) than that for a non-centrosymmetric one (0.736), suggesting the centrosymmetric structure of \ce{FePd2Te2} . Finally, based on its smaller CFOM value, the space group $P2_1/m$ is selected. A structural solution with $P2_1/m$ space group was obtained for \ce{FePd2Te2} using the APEX3 program, and the final refinement was completed with the SHELXL suite of programs.
	
	The powder samples of \ce{FePd2Te2} were also characterized by a Bruker D8 Advance X-ray diffractometer. The elemental composition of single crystals were examined with energy dispersive x-ray spectroscopy (EDS, Oxford X-Max 50), the results are consistent with the chemical formula. Magnetization and electrical transport measurements on bulk samples were carried out in Quantum Design MPMS3 and PPMS-14T respectively. The single crystal neutron diffraction experiments were carried out on Xingzhi cold neutron triple-axis spectrometer at the China Advanced Research Reactor (CARR)\cite{Xingzhi}. The incident neutron energy is fixed at 16~meV with a neutron velocity selector to remove higher-order neutrons.

\subsection{STM measurements.}
	The samples were cleaved at room temperature in ultrahigh vacuum at a base pressure of 2$\times$10$^{-10}$ Torr and directly transferred to the STM system (PanScan Freedom, RHK). Chemically etched Pt-Ir tips were used for STM measurement in constant-current mode at around 10~K. The tips were calibrated on a clean Ag(111) surface. Gwyddion was used for STM data analysis.

\subsection{Exfoliation and Nanoflake Characterization.}
	
Mechanical exfoliation of bulk \ce{FePd2Te2} was achieved using scotch tape and transferred onto a 285~nm \ce{SiO2} covered Si substrate. The atomic force microscopy (AFM) measurements were performed using a commercial Park Systems (Park NX10) with the commercial AFM probes (AC160TS, Nanosensors). The blue tape was used to peel off the samples in a thin layer. Electrodes were deposited on top of the sample using magnetron sputtering. The electrical transport measurements of micro-nano devices were carried out in a Quantum Design PPMS.

\subsection{Density Functional Theory Calculations.}
Density functional theory (DFT) calculations in this study were conducted using the Vienna ab initio simulation package (VASP)\cite{T1,T2}. The projector augmented wave (PAW) potential and the Perdew–Burke–Ernzerhof (PBE) functional within the generalized gradient approximation (GGA) were employed\cite{T3,T4}. The kinetic energy cutoff was set to 500~eV. Ionic relaxations were performed until the forces on the ions were less than 0.02~eV/\AA. For structural optimization, the Monkhorst–Pack method was used to select the $k$-point meshes with dimensions of 3$\times$9$\times$1\cite{T5}. The interlayer interaction energy was calculated by subtracting twice the energy of a single-layer \ce{FePd2Te2} system from the total energy of a bilayer \ce{FePd2Te2} system and then divided by the contact area. A vacuum layer exceeding 10~\AA was set, and the in-plane lattice parameters were allowed to relax freely.

\section{Results and Discussion}
	
\begin{figure}[tbp]
	\centerline{\includegraphics[scale=0.46]{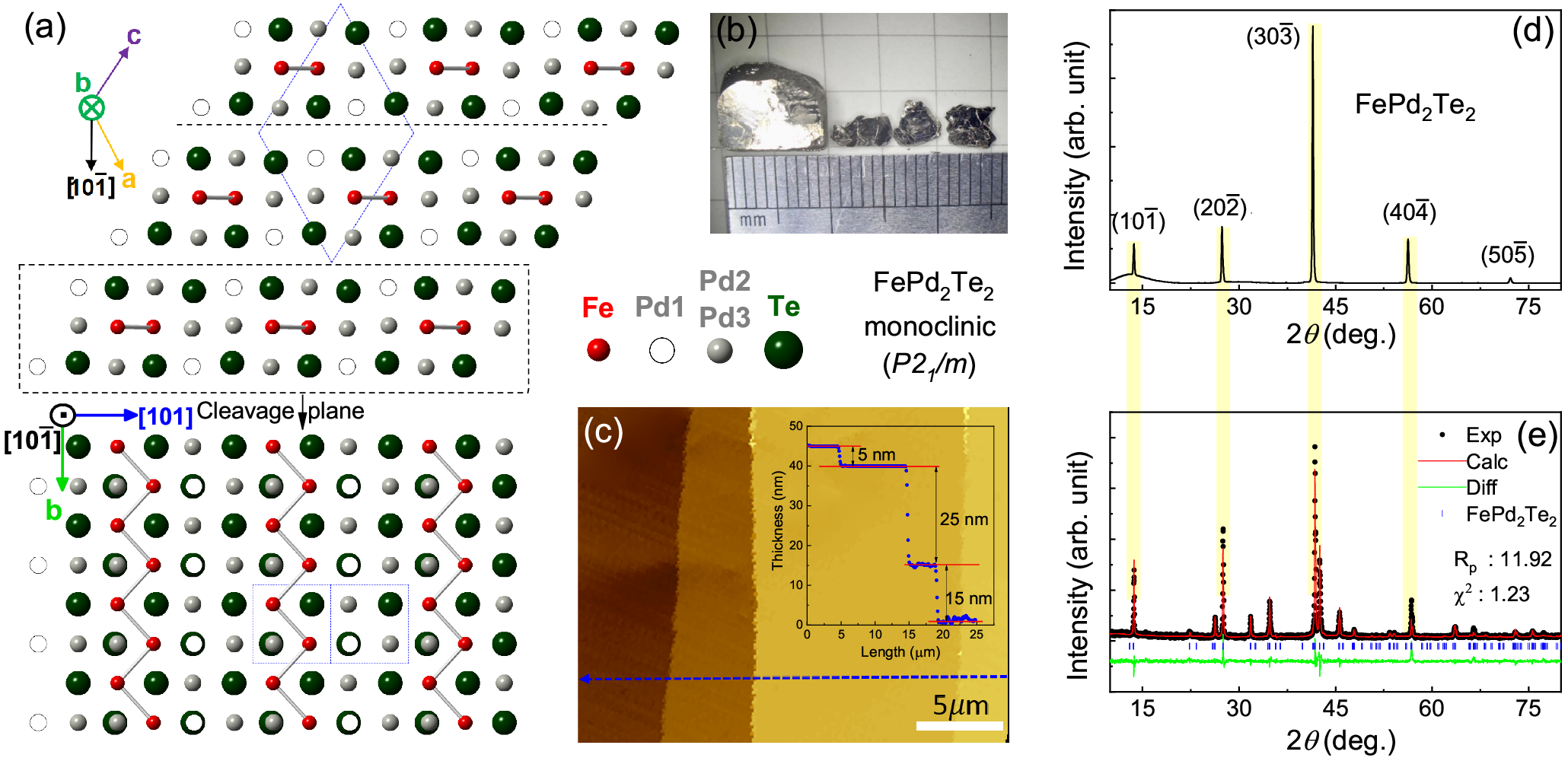}} \vspace*{-0.3cm}
	\caption{Structural characterization of \ce{FePd2Te2}. (a) The crystal structure of \ce{FePd2Te2} shows a layered structure along [$10\overline{1}$] direction. The blue dotted line shows the unit cell. The crystal can be easily exfoliated between the Pd-Te layers indicated by the black dashed line and its cleavage plane is the ($10\overline{1}$) plane which is shown in the bottom. The Fe zigzag-chains lie in this plane and extends along the $b$-axis. (b) A photo of typical \ce{FePd2Te2} single crystals. (c) Atomic force microscopy images and height profile step of a \ce{FePd2Te2} nano-flake demonstrate its 2D layered structure. (d) XRD patterns from the cleavage plane of a single crystal.  (e) Powder XRD pattern of \ce{FePd2Te2}, samples were crushed from single crystals. The yellow bars mark the strongly enhanced $(H0\overline{H})$ Bragg peaks.}
\end{figure}

\begin{table*}
	\centering
	\caption{Atomic Coordinates, Occupancy, and Anisotropic Thermal Parameters for  \ce{FePd_{2.08}Te2}.}
	\begin{tabular}{cccccc}
		\hline\hline
	atom & Wyckoff & occupancy & x  & y & z  \\
	Te1 & 2e & 1 & 0.2293(6) & 0.75 & 0.0045(6)  \\
	Pd1 & 2e & 0.0784 & 0.5274(6) &  0.75 & 0.7232(0) \\
	Te2 & 2e & 1 & 0.7182(5) & 0.75 & 0.5183(1) \\
	Pd2 & 2e & 1 & 0.0388(2) & 0.75 & 0.2312(5) \\
	Pd3 & 2e & 1 & 0.6267(9) & 0.75 & 0.1397(7) \\
	Fe1 & 2e & 1 & 0.1167(1) & 0.75 & 0.6150(3)  \\
	\\
	\hline
	\\
	atom &  $U_{\rm 11}$&  $U_{\rm 22}$&  $U_{\rm 33}$&  $U_{\rm 13}$ & $U_{\rm 12}=$ $U_{\rm 23}$\\
	Te1 & 0.0186(3) & 0.0169(3)	& 0.0176(3) & 0.0088(2) & 0 \\
	Pd1 & 0.0440(6) & 0.0160(4) & 0.0210(4) & 0.0150(4) & 0 \\
	Te2 & 0.0186(3) & 0.0178(3) & 0.0197(3) & 0.0076(2) & 0 \\
	Pd2 & 0.0220(4) & 0.0229(4) & 0.0235(4) & 0.0115(3)	& 0 \\
	Pd3 & 0.0200(4) & 0.0212(4) & 0.0239(4) & 0.0083(3)	& 0 \\
	Fe1 & 0.0209(6) & 0.0173(6) & 0.0170(6) & 0.0096(5)	& 0 \\
	\hline\hline.
	\end{tabular}
	\label{1}
\end{table*}

The discovery of \ce{FePd2Te2} emerged from our search for new compounds in Pd-Fe-Te ternary phase. Fig. 1(b) shows the large single crystal with size up to centimeters could be grown, which is only limited by the diameter of crucibles. Its crystal structure was solved through the refinement of single crystal XRD described in the experimental section and identified as a new prototype structure. We have also checked possible structural similarities with other known compounds such as superconductor \ce{Pd3Te2}\cite{Pd3Te2}, but there are fundamental differences including the crystal system, space group and lattice constants. As shown in Fig. 1(a), \ce{FePd2Te2} adopts the monoclinic symmetry with space group $P2_1/m$ (No.11). The unit cell is illustrated using blue dotted line with the refined lattice parameters a=7.5024(5) \AA, b=3.9534(2) \AA, c=7.7366(7) \AA, $\alpha$=$\gamma$=90$^{\circ}$, $\beta$=118.15(0)$^{\circ}$. Detailed crystallographic and refinement data is shown in Table 1, Table S1 and Fig. S1 in the supplementary material. The accurate chemical formula of this compound should be FePd$_{2.08}$Te$_2$. Since one of the crystallographic positions of Pd (named Pd1) has a negligible occupancy of less than 8\% which is marked by hollow circles in Fig. 1(a), we will use the chemical formula \ce{FePd2Te2} in the following description for simplicity. All other atoms have full occupancy as revealed from XRD refinement. The first structural feature of \ce{FePd2Te2} is that all Fe atoms form infinite zigzag chains along the $b$-axis. The nearest inter-chain distance is about 7.50$\AA$ while the intra-chain nearest Fe-Fe distance is only 2.68$\AA$, exhibiting a quasi-one-dimensional spin system feature since Fe should mainly contribute to the magnetism in this material. Secondly, we find the as-grown crystals are thin plate-like with typical 2D material feature. This is supported by the atomic force microscopy images shown in Fig. 1(c). Through mechanical exfoliation of \ce{FePd2Te2} crystals using Scotch tape, nano-sheet with thickness of $\sim$5~nm could be obtained (Fig. S2). By performing XRD on the cleavage plane, the Bragg peaks are indexed as ($H0\overline{H}$) as shown in Fig. 1(d). Therefore, the cleavage plane is ($10\overline{1}$) plane which is illustrated in the bottom of Fig. 1(a). It should be noted that there is a small intersection angle (less than 2$^{\circ}$) between [$10\overline{1}$] and the normal vector of ($10\overline{1}$) cleavage plane due to the slight difference between $a$- and $c$-lattice constants. In the following magnetic characterization of \ce{FePd2Te2}, we neglect this tiny difference since the experimental error in the field orientation is comparable to this difference. The Fe zigzag chains lie in the cleavage plane. In addition, the single crystals were crushed into powders to carry out powder XRD. Fig. 1(e) shows the diffraction data which can be well fitted by the crystal structure solved from single crystal XRD. It should be mentioned that the intensities of ($H0\overline{H}$) Bragg peaks are much stronger than the calculated intensities, so preferred crystal orientation along ($H0\overline{H}$) must be used in the Rietveld refinement to achieve the best fit. This further provides evidence for the 2D nature of \ce{FePd2Te2}. 

DFT calculation further reveals that \ce{FePd2Te2} should be a kind of 2D material. The interlayer bonds of \ce{FePd2Te2} are mainly Pd2-Te bonds as shown in Fig. 1(a), which has an energy of 0.31~eV/bond. This value is larger than the energy of van der Waals bond but much lower than that of common covalent bond. It could be classified as the covalent-like quasi-bonding as in many other famous 2D materials such as black phosphorus\cite{Hu}, \ce{PtS2}\cite{PtS2} and \ce{PtSe2}\cite{PtSe2}. Especially the Pd2-Te bond has a very low density in the cleavage plane comparing to that of many other 2D materials. So the final calculation result shows that the cleavage energy for \ce{FePd2Te2} is 0.32~J/m$^2$, which is comparable with \ce{CrI3} (0.28 to 0.30~J/m$^2$) and smaller than experimentally estimated value for graphite (0.36~J/m$^2$). Although there are some additional Pd1-Te bonds between layers, the negligible occupancy of Pd1 would not much affect the cleavage energy.

\begin{figure}[tbp]
	\centerline{\includegraphics[scale=0.62]{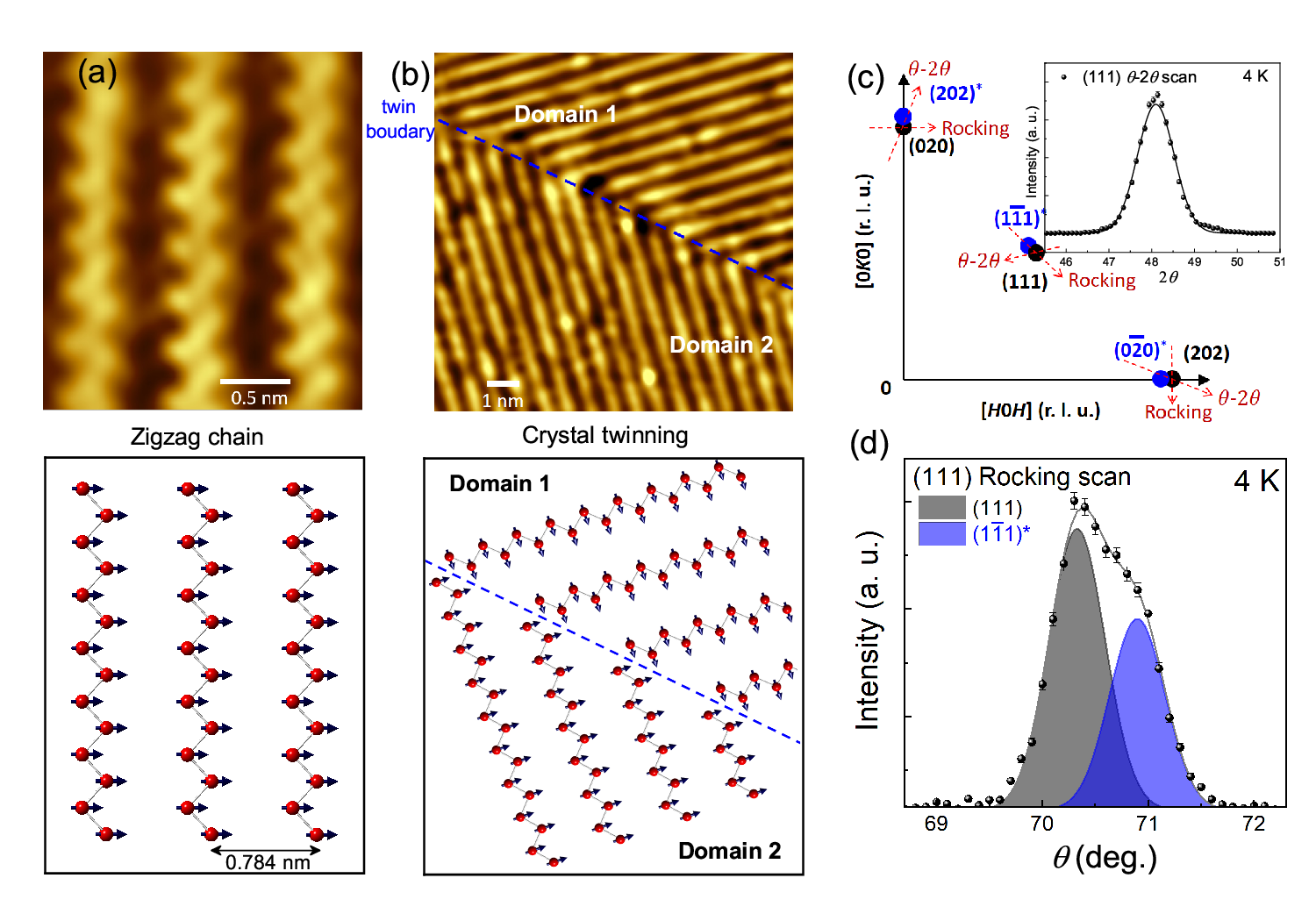}} \vspace*{-0.3cm}
	\caption{Crystal twinning in \ce{FePd2Te2}. (a) Visualization of Fe zigzag chains by STM. Structure of Fe chains determined from XRD is shown in the bottom. (b) Visualization of crystal twinning in the cleavage plane. The blue dashed line indicates the twin boundary. Schematic of the magnetic and crystal structure of Fe-chains in twinning \ce{FePd2Te2} is shown in the bottom. (c) Illustration of the positions for some nuclear Bragg peaks and the directions for different scans in the reciprocal lattice of a \ce{FePd2Te2} twinning crystal. The black and blue points indicate the major domain and twinning domain respectively. The inset shows a $\theta$-$2\theta$ scan on (111) peak using neutron diffraction. (d) A neutron rocking scan on the (111) Bragg peak.}
\end{figure}

STM and neutron diffraction measurements reveal crystal twinning effect in \ce{FePd2Te2}. First of all, the Fe zigzag chains are clearly visualized by atomic-resolution topography STM image of the cleavage plane shown in Fig. 2(a) and (b). There are some defect-like features along the Fe-chain which may possibly be caused by the partial occupancy of Pd1 atoms. From XRD refinement, the nearest Fe-Fe-Fe bonding angle is about 95$^{\circ}$ and the adjacent two Fe-chains in the cleavage plane have a distance of 7.84\AA. These parameters are consistent with the observation from STM. Furthermore, the STM image captures the existence of crystal twinning in \ce{FePd2Te2}. As shown in Fig. 2(b), two kinds of domains have a twin boundary along the [111] crystal orientation which is indicated by the blue dashed line. Unlike some other twinning crystals where arbitrary twinning angles can be random distributed, the twinning angle of \ce{FePd2Te2} is fixed at 90$^{\circ}$. The crystal twinning has an intriguing result which is making the Fe-chains abruptly turn 90$^{\circ}$ in the domain boundary. The twinning crystal structure of the Fe-chains is also illustrated in the bottom of Fig. 2(b). This crystal twinning effect can also be detected by single crystal neutron diffraction performed on Xingzhi. As illustrated in Fig. 2(c), for such a twinning crystal structure, the position of Bragg peak (111) from the major domain (black points) is close to (1$\overline{1}$1)* peak from the twinning domain (blue points) in the reciprocal space. In principal, the $\theta$-$2\theta$ scan on (111) would only detect a single peak while the rocking scan would capture double peaks because of different scanning directions (the result would be contrary for scans on (020) and (202)). This is indeed consistent with the neutron experimental results as shown in Figs. 2(c), (d) and Fig. S3. In addition, the integrated intensity ratio of (111) to (1$\overline{1}$1)* in the rocking scan is 6:4 which reflects the volume ratio of two domains. This result has been reproduced in another sample and is consistent with the following susceptibility analysis which means that nearly 60\% of the crystal domains belong to the major domain.

The crystal twinning effect in \ce{FePd2Te2} may possibly be driven by the internal stress. Fig. S4 illustrates the crystal structure of splicing two 90$^{\circ}$ twinning domains in the cleavage plane. It can be seen that the arrangement of Te atoms in the cleavage plane has a nearly square-lattice four-fold symmetric feature. A 90$^{\circ}$ turn of the domain would not affect the symmetry of Te-lattice, which is compatible with two different domains. On the other hand, the one-dimensional Fe chains may create internal stress along one direction, while a 90$^{\circ}$ turn of the chains might naturally release the internal stress during the crystal growth. This is a possible origin for the crystal twinning in \ce{FePd2Te2}.

Magnetization measurements were carried out on \ce{FePd2Te2} single crystal to investigate its magnetic properties. Fig. 3(a) shows the temperature dependent susceptibility $\chi(T)$ curves. For either applying field parallel or perpendicular to the cleavage plane, a ferromagnetic (FM) transition could be observed at Curie temperature $T_C$=183~K. This is further confirmed by the well-defined peak appears in the AC susceptibility (ACS) $\chi^{''}(T)$ at the same temperature (Fig. 3(b)). ACS also has another small peak at around $T^{*}$=100~K. We do not observe any anomaly from specific heat data at $T^*$ (Fig. S5). This anomaly and the drop of ZFC susceptibility in Fig. 3(a) at low temperatures are similar as that in Fe$_3$GeTe$_2$, possibly due to some change and movement in the magnetic domains\cite{Yi,Kong,CP_Fe3,Guo}. In addition, the long-range magnetic order developed below $T_C$ is also confirmed by neutron diffraction data shown in Fig. 3(c). Two typical nuclear Bragg peaks (111) and ($20\overline{2}$) exhibit significant magnetic intensity contributions below $T_C$. The isothermal magnetization at 2~K and 75~K under different field directions are shown in Figs. 3(d) and (e). Notable anisotropic magnetization behavior could be observed. When the field is applied parallel to the cleavage plane (H$\Vert$($10\overline{1}$)), the saturation field is much smaller than that when H$\perp$($10\overline{1}$), indicating that \ce{FePd2Te2} is an easy-plane ferromagnet.

\begin{figure}[tbp]
	\centerline{\includegraphics[scale=0.47]{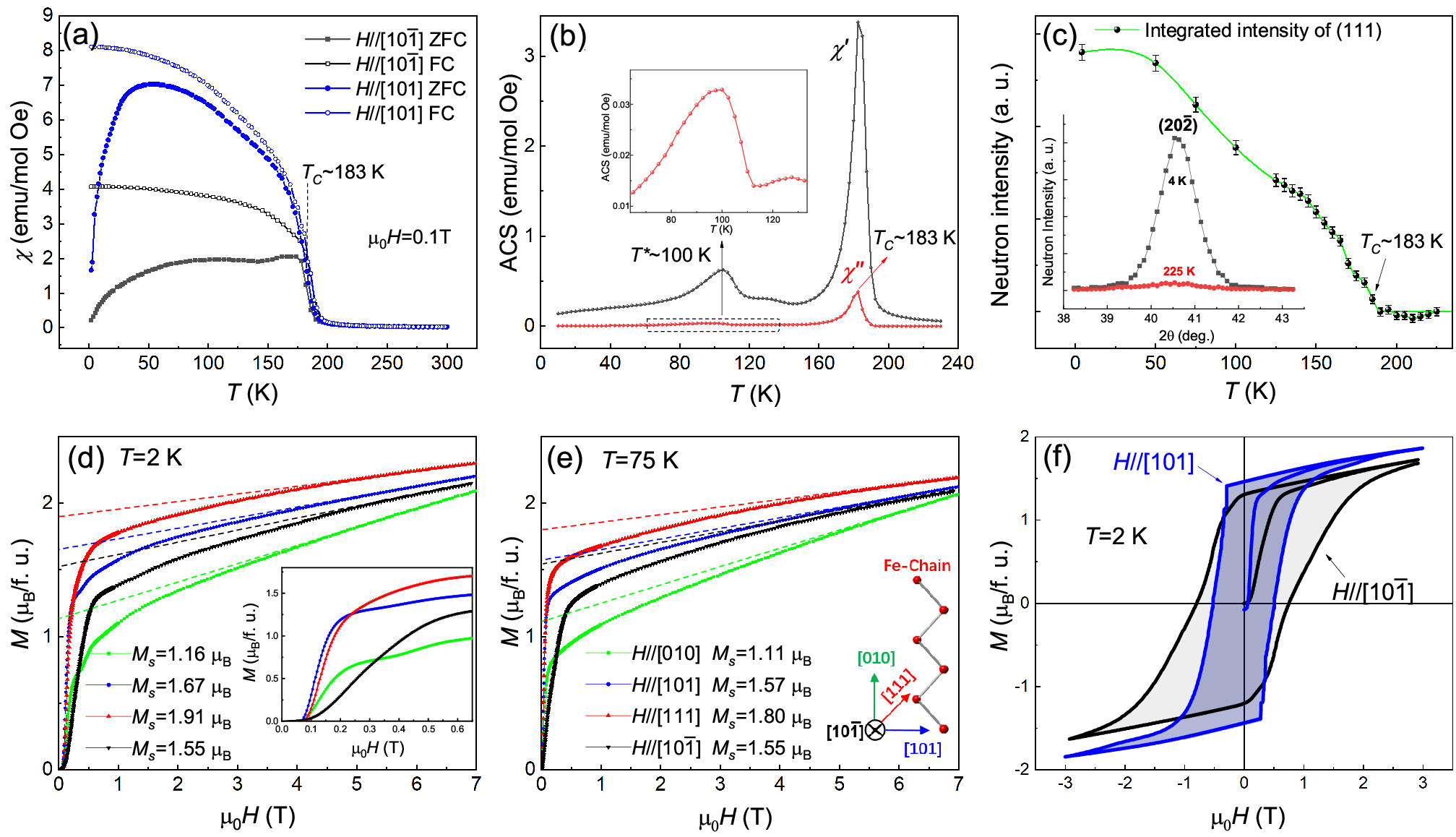}} \vspace*{-0.3cm}
	\caption{Magnetic properties of \ce{FePd2Te2}. (a) Temperature dependent magnetic susceptibility of \ce{FePd2Te2} single crystal at $\mu_{0}H=$ 0.1~T. (b) AC magnetic susceptibility measured under $H\Vert [101]$. The inset shows an enlarged view of $\chi_{ac}^{''}(T)$ around 100~K. (c) The integrated intensity of Bragg peak (111) as the order parameter of magnetic transition measured by neutron diffraction. The inset shows another peak ($20\overline{2}$) measured below and above T$_C$. (d,e) Isothermal magnetization along different crystal orientations with respect to the Fe zigzag chain shown in the inset of (e). The inset of (d) shows an enlarged view on the low field data. (f) Magnetic hysteresis at 2~K under maximum field of 3~T.}
\end{figure}	

\begin{table*}
	\centering
	\caption{Curie-Weiss Temperature, Effective Moment, and Rhodes-Wohlfarth Ratio for \ce{FePd2Te2}.}
	\begin{tabular}{cccc}
		\hline\hline
		Data for fitting & $\theta_{CW}$ & Effective Moment & RWR  \\
		$H\Vert [10\overline{1}]$ & 173.2~K & 5.07 $\mu_{B}$/Fe & 2.688 \\
		$H\Vert [101]$ & 172.0~K & 5.24 $\mu_{B}$/Fe & 2.595 \\
		$H\Vert [010]$ & 168.3~K & 5.38 $\mu_{B}$/Fe	& 3.855 \\ 
		$H\Vert [111]$ & 173.5~K & 5.17 $\mu_{B}$/Fe & 2.233 \\
		\hline\hline
	\end{tabular}
	\label{2}
\end{table*}

Furthermore, \ce{FePd2Te2} also exhibits strong in-plane magnetic anisotropy when applying fields along different in-plane directions. The inset of Fig. 3(e) has shown three major crystal-axes in the cleavage plane. These are [010] along the extending direction of Fe-chain ($b$-axis), [101] which is perpendicular to the chain and [111] that is rotating about 45$^{\circ}$ from [010]. The laue X-ray image on ($10\overline{1}$) cleavage plane shown in Fig. S6 could enable us to distinguish these three axes, although there is some interference from crystal twinning. From the inset of Fig. 3(d), the fastest magnetization saturation is identified to occur under H$\Vert$[101], which implies that the easy-axis should be along [101]. However, in Fig. 3(d) and (e), the determined saturation moment $M_s$ has the largest value when H$\Vert$[111]. It seems that [111] may also be the possible magnetization easy-axis. We argue that the easy-axis of \ce{FePd2Te2} should still be [101] and the largest $M_s$ value for H$\Vert$[111] could be well explained combined the uniaxial magnetic anisotropy and crystal twinning effect in \ce{FePd2Te2}. Namely, \ce{FePd2Te2} has strong uniaxial magnetic anisotropy with the moment pointing to [101]-axis. When encountering crystal twinning, that two kinds of domains coexist with intersection angle of 90$^{\circ}$, the magnetic structure should be like the model illustrated in Fig. 2(b). In this picture, the so-called $M_s^{[101]}$ should count moments only from the major domains, while $M_s^{[010]}$ should count ones only from the twinning domain. Namely $M_s^{[010]}$ is actually $M_s^{[101]^*}$ in the twinning domain. For such an uniaxial anisotropy, $M_s^{[111]}$ should be the sum of projections from both $M_s^{[101]}$ and $M_s^{[010]}$. Namely, it should satisfy the following equation:

\begin{equation}
	M_s^{[111]}=\cos45^{\circ}(M_s^{[101]}+M_s^{[010]})
\end{equation} 

The calculated $M_s^{[111]}$ based this equation is consistent with the experimental $M_s^{[111]}$ for both 2~K and 75~K. The maximum error is only 5$\%$ and similar results were reproduced in several crystals, which provides evidence for the magnetic structures depicted in Fig. 2(b). Large deviation from equation (1) only occurs for magnetization at temperatures close to $T_C$ possibly due to large thermal fluctuations (Fig. S7). In addition, one can use the value of $M_s^{[101]}$ /$M_s^{[010]}$ to estimate the volume ratio of these two kinds of domains, which is also close to $\sim$6:4 and reaches perfect agreement with that determined from neutron diffraction. This further validates the magnetic anisotropy and spin configuration model proposed by us. These consistent results strongly suggest the intimate relation between microstructure twinning and macro-magnetism in \ce{FePd2Te2}. According to STM result, one single domain has typical scales from 10~nm to 100~nm, thus an  intriguing spin texture are likely naturally formed in the cleavage plane of \ce{FePd2Te2} which may stimulate further researches.

Figure. 3(f) shows the large magnetic hysteresis when the field is applied either in-plane ($H\Vert$[101]) or out-of-plane ($H\Vert [10\overline{1}]$). Both the large coercive field $H_{c}$ (0.77~T for $H\Vert [10\overline{1}]$ and $\sim$0.51~T for $H\Vert [101]$) and the high magnetic remanence to saturated magnetization ratio $M_{R}$/$M_{s}$ ($\sim$0.94) shows that \ce{FePd2Te2} is a hard magnet which is preferred for spintronic application\cite{Fe3_hard}. Furthermore, through the Curie-Weiss (CW) fit of high-temperature paramagnetic susceptibility, the CW temperature $\theta_{CW}$ and the effective magnetic moment per Fe ion could be obtained and shown in Table 2. Although there are slight differences for different field directions, generally the values of $\theta_{CW}$ are consistent with the ferromagnetic interaction below $T_C$. The effective moment per Fe is a bit larger than the theoretical value considering only the spin angular momentum S=2 for Fe$^{2+}$, which suggests the unquenched orbital contribution. Combined with the saturation moment M$_s$ obtained from linear extrapolation of high-field $M(H)$ curve, we could calculate the Rhodes-Wohlfarth ratio\cite{RWR} (RWR) as shown in Table 2. If considering the previous anisotropy and crystal twinning analysis, $M_s^{[101]}$ should be 2.83$\mu_{B}$ for a single domain crystal which yields RWR=1.53. Since RWR=1 corresponds to the purely localized magnetism, the larger RWR values in our samples and the fact that \ce{FePd2Te2} is metallic suggest that itinerant magnetism is likely present in this compound\cite{RWR,RWR_Fe5}.

\begin{figure}[tbp]
	\centerline{\includegraphics[scale=0.5]{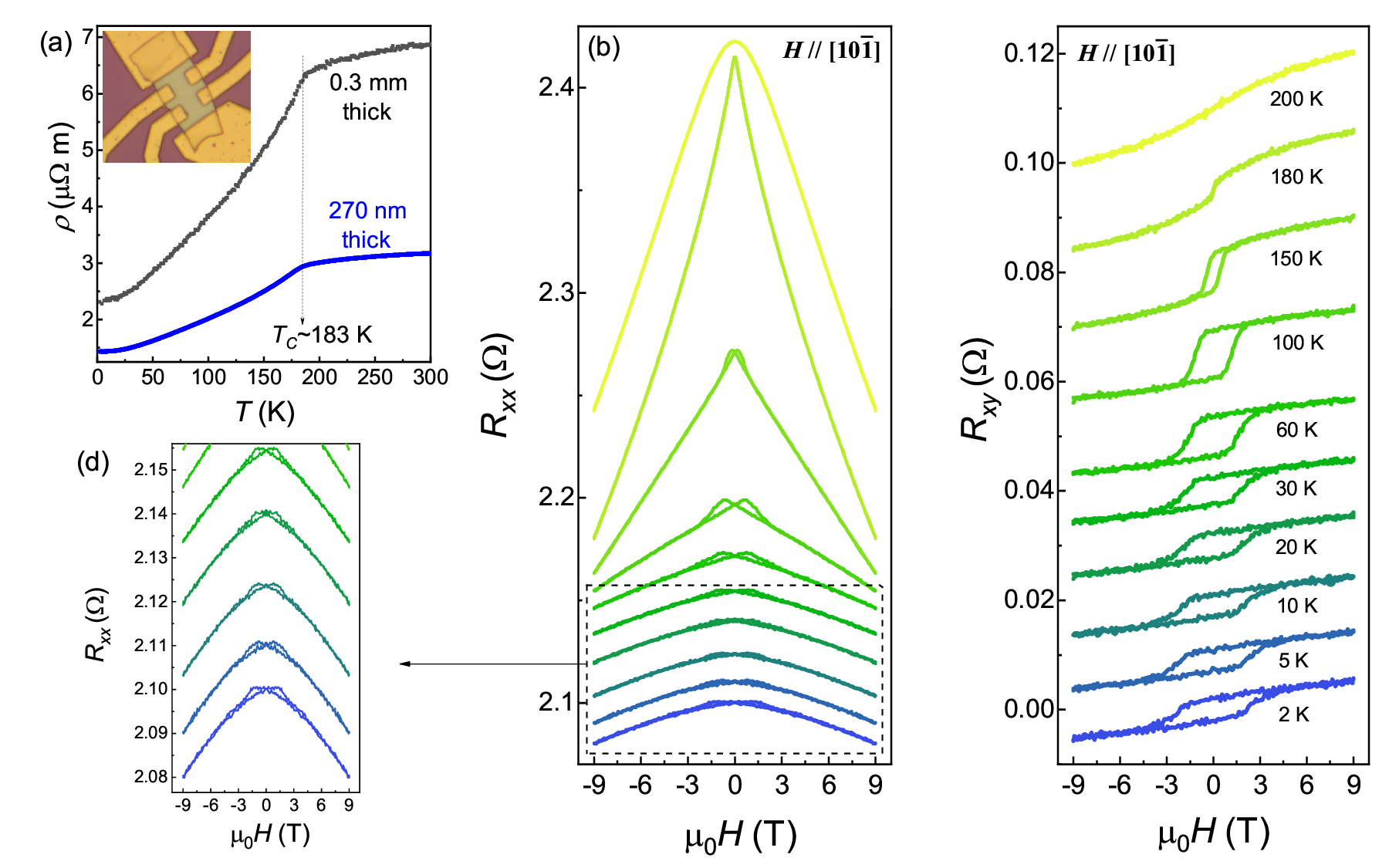}} \vspace*{-0.3cm}
	\caption{Transport properties of \ce{FePd2Te2}. (a) Temperature-dependent longitudinal resistivity of \ce{FePd2Te2} for crystals with different thickness. The inset shows the optical micrograph of the 270~nm-thick device which is used for the following MR and Hall effect measurements. (b) The magnetic field dependence of longitudinal resistance $R_{xx}$ at different temperatures under $H\Vert [10\overline{1}]$. (c) The Hall resistance $R_{xy}$ measured at different temperatures. (d) An enlarged view of the low temperature MR.}
	\label{figb4}
\end{figure}

\begin{figure}[bp]
	\centerline{\includegraphics[scale=0.35]{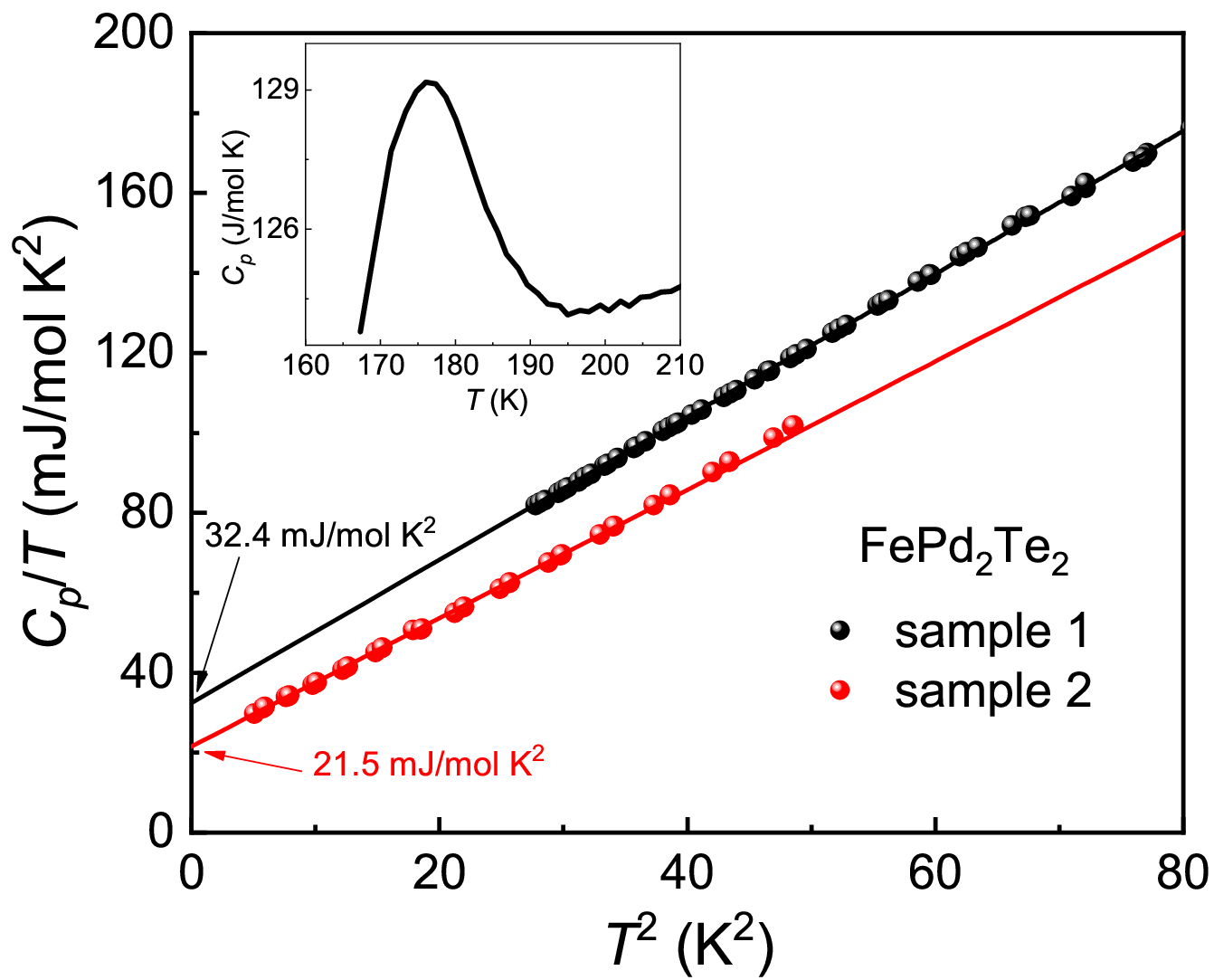}} 
	\vspace*{-0.3cm}
	\caption{Results of specific heat $C_P(T)$ measurements. A representative plot of $C_P(T)/T$ vs $T^2$ at low temperature is shown. The inset shows the $C_P(T)$ data at around Curie temperature.}
	\label{SH}
\end{figure}

Next, we present the electrical and magneto-transport results. As shown in Fig. 4(a), for both \ce{FePd2Te2} bulk and a 270~nm thin flake, the temperature dependent resistivity exhibits a kink at $T_C$. This is a typical feature for FM ordering caused by the reduced spin scattering in the FM state. Figs. 4(b)-(d) display the magnetoresistance (MR) and Hall effect of the 270~nm-thick device with magnetic field applied perpendicular to the cleavage plane. Although \ce{FePd2Te2} has a strong in-plane magnetic anisotropy, sufficiently strong out-of-plane field could also polarize it along this direction. Firstly, the existence of robust FM order in the flake can be seen from both the strong anomalous Hall signal and the butterfly-shaped hysteresis in the low-field MR. Moreover the $R_{xy}$ loop exhibits an extremely large coercivity of 2.1~T at 2~K. For comparison, the thick bulk \ce{FePd2Te2} crystal only has a coercivity of 0.77~T as shown in Fig. 3(f). This enhanced hard magnetic behavior due to the reduced thickness is also observed in other 2D magnets such as \ce{Fe3GeTe2}\cite{Fe3_hard} and \ce{Fe3GaTe2}\cite{Ga_hard}. It should be noted that the coercivity of \ce{FePd2Te2} is much larger than both \ce{Fe3GeTe2} and \ce{Fe3GaTe2} at similar thickness and temperature\cite{Fe3_hard,Ga_hard}. This hard magnetic property could be advantageous for spintronic applications. We should also mention that \ce{FePd2Te2} is twinned and how the crystal twinning may affect coercive field is currently not clear. In addition, $R_{xy}$ shows a linear $H$-dependent behavior under field higher than 6~T. Then we can use the slope of high-field $\rho_{xy}$  to estimate the carrier density of \ce{FePd2Te2} based on a simple single-band analysis. The results are shown in Fig. S8 and hole-like charge carries are identified. This value is likely impacted by the existence of multi-carrier transport and further conclusion could not be drawn without detailed knowledge of the Fermi surface. On the other hand, there is a linear and non-saturating negative MR at high field region up to 9~T as seen from Figs. 4(b) and (d). The negative MR should be caused by suppressing fluctuating magnetic moments with increasing field, which reaches the highest value of $\sim$10$\%$ at 180~K near the transition temperature. The origin of H-linear MR may be elusive. Possible explanations include field suppression of magnons\cite{MR_magnon}, the field-induced modification to the phase-space volume owing to the nonzero Berry curvature\cite{MR_berry} and Kondo effect\cite{MR_kondo}. This is an interesting topic for future investigations.

Figure. 5 shows the results of specific heat $C_{p}(T)$ measurements. In the high-temperature region, a jump corresponding to the FM transition is observed near $T_C$. At low temperatures, the data could be well fitted by the expression $C_{p}(T)/T=\gamma+\beta T^2$ as shown in Fig. 5. Then the electronic specific heat coefficient $\gamma$ could be obtained. We found that $\gamma$ is sample dependent. For measurements on several different crystals, the values are between 32.4 and 21.5 mJ mol$^{-1}$ K$^{-2}$/Fe. The large $\gamma$ values indicate a strong electron correlation in this material. We notice this value is comparable with other strongly correlated metals such as \ce{Fe3GeTe2} (45 mJ mol$^{-1}$ K$^{-2}$/Fe)\cite{Fe3_corre1,Fe3_corre2,Fe3_corre3} and \ce{SrRuO3} (29 mJ mol$^{-1}$ K$^{-2}$)\cite{SrRuO3}. Additionally the coefficient $\beta $ could also be obtained from the above fit and the Debye temperature is estimated to be $\Theta_{D} $ = 176~K for sample 1 and 182~K for sample 2.

Finally, we would like to have some discussions on the properties of \ce{FePd2Te2}. Besides its air-stability, high Curie temperature and hard magnetic behavior, \ce{FePd2Te2} serves as a fascinating and unique example of 2D magnets at least from four aspects. Firstly, large centimeter-size single crystals could be grown and their size seems to be only limited by the crucible size. This advantage not only opens broad prospects for device applications that required large-scale integration\cite{Liu}, but also allows more experimental techniques such as inelastic neutron scattering to further investigate its physical properties. Secondly, due to the existence of well separated Fe-chains, \ce{FePd2Te2} is a rare example for studying one-dimensional magnetism in 2D material. It is also a suitable material for the research on competition between itinerant and local moment magnetism, as well as Kondo physics, just as the recent discoveries in \ce{Fe3GeTe2} via inelastic neutron scattering.\cite{Bao}

Thirdly, \ce{FePd2Te2} is also a rarely seen 2D magnetic metal with strong in-plane uniaxial magnetic anisotropy. Most 2D magnets studied so far have perpendicular anisotropy, some recently discovered easy-plane magnets such as CrSBr\cite{CrSBr1} and DyOCl\cite{DyOCl} are antiferromagnetic semiconductor or insulators. The in-plane anisotropy of \ce{FePd2Te2} may originate from the strong spin-orbital coupling effect as the existence of many heavy atoms such as Pd and Te. The strong in-plane anisotropy combined with the metallic nature may open up new opportunities for spintronic studies. At last, the crystal twinning effect combined with the in-plane uniaxial magnetic anisotropy naturally created exceptional spin textures as shown in Fig. 3(f), suggesting new possibilities for exploring quantum magnetic phases as that in the CrSBr spin texture created by electron beam\cite{CrSBr_texture}. In addition, further investigations on how the growth or annealing conditions may influence the crystal twinning or even completely detwin \ce{FePd2Te2} may simultaneously find the method for controlling and engineering the spin texture in \ce{FePd2Te2}, paving the ways for new discoveries in 2D magnetism.

\section{Conclusion}

	In summary, we have discovered \ce{FePd2Te2} as an air-stable correlated 2D magnet with Curie temperature of 183~K and hard magnetic properties. Its one-dimensional Fe zigzag chains, uniaxial in-plane magnetic anisotropy and exceptional spin texture created by crystal twinning, make \ce{FePd2Te2} a unique material platform for studying both low-dimensional magnetism and spintronic applications. 

\section{Associated Content}

\subsection{Supporting Information}

Additional single crystal X-ray refinement and neutron diffraction data; atomic force microscopy image on nano-flakes; single crystal Laue X-ray image; illustration of twinning crystal structure; Hall coefficient analysis; additional magnetization and specific heat data.

\subsection{Notes}

The authors declare no competing financial interest.

\acknowledgement

We thank Mr. Youting Song for the help in single crystal XRD measurement and analysis. This work is supported by the National Natural Science Foundation of China (No. 12074426, No. 12004426, No. 11227906), the National Key R\&D Program of China (MOST) (Grant No. 2023YFA1406500), the Fundamental Research Funds for the Central Universities, and the Research Funds of Renmin University of China (Grants No. 22XNKJ40 and No. 21XNLG27), NSAF (Grant No. U2030106). Y. Y. Geng was supported by the Outstanding Innovative Talents Cultivation Funded Programs 2023 of Renmin University of China.

	$^{\dag}$ These authors contributed equally to this work.

\begin{figure}
	%\caption{}%Õâžö±êÌâÒÑŸ­È¡Ïû
	\begin{center}
		TOC
	\end{center}
	\centerline{\includegraphics[scale=1.5]{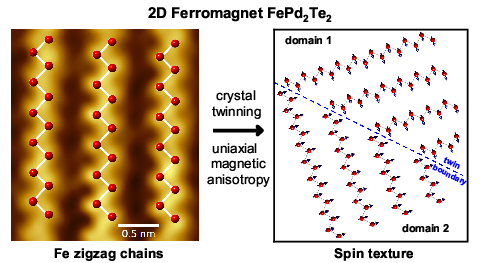}} \vspace*{-0.3cm}
\end{figure}

\end{document}